\documentclass{vgtc}                          % final (conference style)
\graphicspath{{figures/}{pictures/}{images/}{./}} % where to search for the images

\usepackage{gensymb} %for symbol
\usepackage{times}                     % we use Times as the main font
\usepackage{tabu}                      % only used for the table example
\usepackage{booktabs}                  % only used for the table example
\usepackage{lipsum}                    % used to generate placeholder text
\usepackage{mwe}                       % used to generate placeholder figures
\usepackage{cite}           %added for ordered ref %(Sam-0715-#9)

\usepackage{mathptmx}                  % use matching math font

\onlineid{1101}

\vgtccategory{Research}

\vgtcinsertpkg

\title{What Color Scheme is More Effective in Assisting Readers to Locate Information in a Color-Coded Article?}

\author{Ho Yin Ng\thanks{e-mail: hzn5135@psu.edu} %
\and Zeyu He\thanks{e-mail: zmh5268@psu.edu} %
\and Ting-Hao `Kenneth' Huang\thanks{e-mail: txh710@psu.edu}}
\affiliation{\scriptsize Pennsylvania State University}

\abstract{
    Color coding, a technique assigning specific colors to cluster information types, has proven advantages in aiding human cognitive activities, especially reading and comprehension.
The rise of Large Language Models (LLMs) has streamlined document coding, enabling simple automatic text labeling with various schemes.
This has the potential to make color-coding more accessible and benefit more users.
However, the impact of color choice on information seeking is understudied. 
We conducted a user study assessing various color schemes' effectiveness in LLM-coded text documents, standardizing contrast ratios to approximately 5.55:1 across schemes. 
Participants performed timed information-seeking tasks in color-coded scholarly abstracts. 
Results showed non-analogous and yellow-inclusive color schemes improved performance, with the latter also being more preferred by participants. These findings can inform better color scheme choices for text annotation.
As LLMs advance document coding, we advocate for more research focusing on the ``color'' aspect of color-coding techniques.
} % end of abstract

\usepackage{xspace}
\usepackage{xcolor}
\usepackage{comment}
\usepackage{multirow}
\usepackage{subcaption}
\usepackage{amssymb} % for symbol

\newcommand{\kenneth}[1]{{\small\textcolor{blue}{\bf [#1 --Ken]}}}
\newcommand{\sam}[1]{{\small\textcolor{orange}{\bf [#1 --Sam]}}}
\newcommand{\steven}[1]{{\small\textcolor{red}{\bf [#1 --Steven]}}}

\newcommand{\eg}{{\it e.g.}\xspace}

\begin{document}


\firstsection{Introduction\label{sec:intro}}

\maketitle

%\kenneth{It's just easier to add label\{\} inside section\{\}. Much easier to manage.}

%% \section{Introduction} %for journal use above \firstsection{..} instead
%\kenneth{TODO Sam: Hmm any idea why ``Package cleveref Error: cleveref must be loaded after amsmath!.'' error keep showing up on Overleaf? It might be fine but do we know how to fix it?} \sam{I fixed it by disabling one symbol package that I used previously and correct the symbol in the Table}

%\kenneth{TODO Sam: Make sure we use PAST TENSE when describing what we did. I fixed a few but please do a pass for fixing tense.} \sam{Done}

%\kenneth{-------------------- KENNETH IS WORKING HERE: BEGINS -------------------}
 %for missing cross-reference (Sam-0706-#2) 
%Color coding is a simple yet effective technique that assigns specific colors to different types of information in an article. 
Color coding is a simple yet powerful technique that assigns specific colors to different types of information within an article, effectively performing a clustering task~\cite{kucher2015text} in the domain of text visualization.
%A classic example commonly taught in U.S. classrooms (grades 6 to 12) is distinguishing between facts and feelings---\textit{facts vs. feelings}---in an essay using color coding~\cite{geigle2014color}. %(Sam-0717-#11)
Despite being simple, color coding is remarkably powerful, helping readers quickly identify key information in an essay and assisting writers in analytically evaluating their own compositions~\cite{olson2007cognitive}.
%It is known that color-coding positively influences reading and comprehension by improving concentration and reading performance~\cite{mirzabeikivisualization,lee2008raising}, reading speed~\cite{wu2003improving}, and aiding word recognition~\cite{sandvold2023revisiting}.
%Numerous studies also concluded that color coding benefits human cognitive activities, such as enhancing memory~\cite{dzulkifli2013influence,pruisner1993color}, freeing cognitive capacities~\cite{folker2005processing}, and organizing information and fostering coherent formation~\cite{marcus1991graphic}.
%(Sam-0717-#11)
%Numerous 
Studies also showed that color coding benefits human cognitive activities~\cite{dzulkifli2013influence,pruisner1993color,folker2005processing,marcus1991graphic}.
Meanwhile, the recent emergence of
% Large Language Models 
LLMs
has transformed the document annotation landscape. 
LLMs enabled effortless labeling of arbitrary text documents with arbitrary label schemes automatically~\cite{gilardi2023chatgpt,tornberg2023chatgpt}.
%Users no longer have to follow the traditional process of developing a text annotation model, which involves curating training data, training or fine-tuning a labeling model, and testing it.
%Instead, they can label a document by simply prompting LLMs.
This development holds significant potential for enhancing the accessibility of color coding, allowing a broader audience to reap its benefits in reading and writing support. 
From a Human-Computer Interaction (HCI) perspective, \textbf{given LLM's advancements in the ``coding'' aspect of color coding, it is opportune to focus more on the ``color'' aspects of this technique.}
We noted a lack of 
%We found that there is a notable absence of 
discourse on the significance of color choice, particularly in supporting information-seeking in text through different color schemes.
%While there have been studies on the psychological and educational benefits of color coding~\cite{folker2005processing,de2009towards,dzulkifli2013influence}, there is a notable absence of discourse on the significance of color choice, especially in supporting information-seeking in text through different color schemes.
%Moreover, most research in visualization relevant to colors was centered around graphical representations like maps~\cite{tominski2008task,lloyd2018metro}, scientific images~\cite{samsel2018art}, or infographics~\cite{yuan2021infocolorizer}, with limited attention given to color-coded documents.
%(Sam-0717-#11)

This paper presents a user study designed to evaluate the efficacy of color schemes in enhancing readers' information-seeking for color-coded documents.
To replicate the scenario of LLM-coded documents, we used GPT-4 to annotate the documents with our specified color schemes.
We designed information-seeking tasks using abstracts from scholarly papers.
%Then we are i
%Inspired by the previous work of studying categorical perception using warm and cool colors~\cite{holmes2017categorical}, 
We developed the main user study with 10 color schemes generated by 4 base colors in 2 distinct color temperatures (warm: red and yellow; cool: blue and green) to investigate how color schemes generated with specific base colors could impact the information-seeking performance in the text document. 
%\sam{Write the result here: 1) what particular color is good, 2) preference} 
% \steven{which is good added. need to add preference from user survey}
%The results (Section \ref{sec-finding}) indicated that the \textbf{non-analogous (mixed color temperature)} color schemes significantly improved overall performance compared to \textbf{analogous} color schemes.  %(Sam-0717-#11)
The results (Section~\ref{sec-finding}) indicated that 
{\em (i)} the \textbf{non-analogous (mixed color temperature)} color schemes significantly improved overall performance compared to \textbf{analogous} color schemes, 
%For response time, \textbf{dichromatic} schemes resulted in shorter response time than \textbf{monochromatic} schemes. %(Sam-0717-#11)
{\em (ii)} \textbf{dichromatic} schemes resulted in shorter response time than \textbf{monochromatic} schemes,
%\textbf{Yellow-inclusive} schemes resulted in a significantly shorter time compared to \textbf{yellow-exclusive} schemes and were more preferred by most participants. %(Sam-0717-#11)
{\em (iii)} \textbf{yellow-inclusive} schemes resulted in shorter response times and were more preferred by most participants,
%Conversely, \textbf{red-inclusive} schemes led to significantly longer response times compared to \textbf{red-exclusive} schemes and were least favored by the participants.%(Sam-0717-#11)
{\em (iv)} \textbf{red-inclusive} schemes led to longer response times and were least favored by the participants.
This paper shows that the choice of color schemes significantly affects readers' performance in seeking information in color-coded documents.

%With LLMs simplifying document coding, we propose further research on the ``color'' aspect of color-coding techniques. --> reviewer suggest to comment out (Sam-0706-#1)
% The study results showed that cool-warm combination and color schemes containing yellow
%We recorded performance data and participants' preferences for the color palettes, 
%We found that 
% the \textbf{sequential} color scheme led to the poorest performance and was the least preferred by all participants.
%\steven{should we briefly explain about three color palettes here? or adding a  ure to show three color palette} \input{Figures/color-palette-brief}
% Conversely, the \textbf{diverging} and \textbf{qualitative} color schemes demonstrated strong performance and were more widely accepted by the participants.

%\input{Figures/color-palette-brief}

\begin{figure*}[t]
    \centering
    \includegraphics[width=\textwidth]%[width=17.5cm]
    {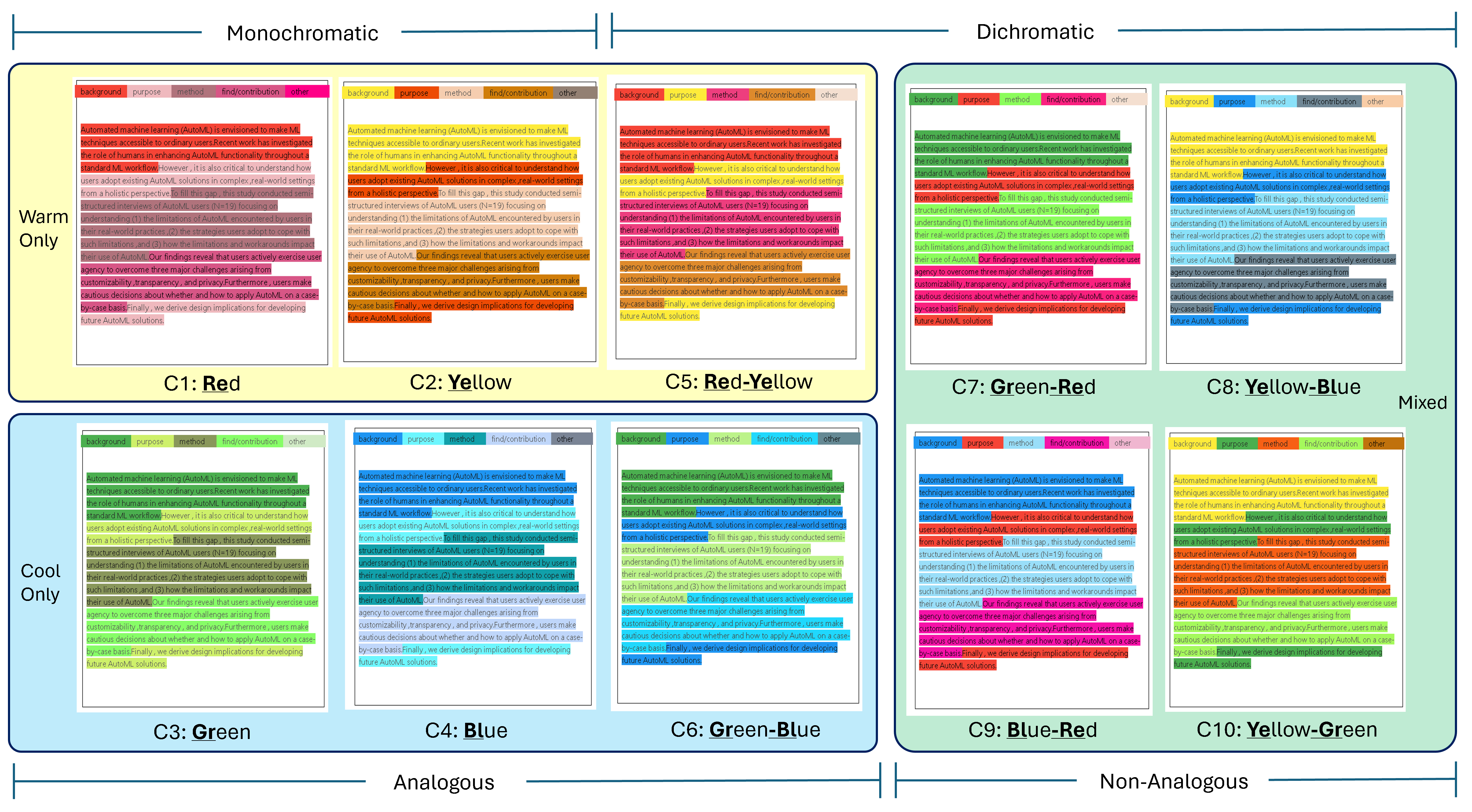}
    \vspace{-.5pc}
    \caption{The 10 color schemes used in our study, generated by combinations of 4 base colors—warm (Red, Yellow) and cool (Green, Blue). \textit{Monochromatic} schemes consist of variations of a single base color; \textit{Dichromatic} schemes are genereated by combining two different base colors; \textit{Analogous} schemes feature base colors with the same or similar hues; \textit{Non-analogous} schemes include contrasting base colors.}%\kenneth{Add more text here to describe how these color schemes were selected.}
    \vspace{-.5pc}
    \label{fig:color-palette-brief}
\end{figure*}

\begin{comment}

\begin{figure*}[t]
    \centering
    \begin{subfigure}[b]{0.3\textwidth}
    \centering
    \includegraphics[width=3.5cm]{Figures/color-palette-bar/sequential.png}
    \caption{Sequential}
    \end{subfigure}
    \begin{subfigure}[b]{0.3\textwidth}
    \centering
    \includegraphics[width=3.5cm]{Figures/color-palette-bar/diverging.png}
    \caption{Diverging}
    \end{subfigure}
    \begin{subfigure}[b]{0.3\textwidth}
    \centering
    \includegraphics[width=3.5cm]{Figures/color-palette-bar/qualitative.png}
    \caption{Qualitative}
    \end{subfigure}
    \vspace{-.5pc}
    \caption{Sequential, Diverging, and Qualitative color schemes used in our study.}
    \vspace{-.5pc}
    \label{fig:color-palette-brief}
\end{figure*}

\end{comment}

\section{Background}
\subsection{Color Coding: Enhancing Learning, Reading, and Information Handling}
Color coding is an efficient technique that utilizes stimuli to enhance information handling~\cite{alluisi1958verbal}.
Dating back to the 1950s, early research explored color's role in visual search tasks for target numbers~\cite{green1956color}. 
Hitt's study %in the same era 
compared coding methods across operator tasks, indicating that color and numeral coding outperformed others~\cite{hitt1961evaluation}.
As mentioned in the Introduction (Section~\ref{sec:intro}), various research supports color coding's advantages in aiding human cognitive activities~\cite{dzulkifli2013influence,pruisner1993color,folker2005processing,marcus1991graphic}, especially reading and comprehension~\cite{mirzabeikivisualization,lee2008raising,wu2003improving,sandvold2023revisiting}.
Recognizing these benefits, color coding has gained popularity as a tool in classrooms, supporting student learning and reading.
It enables text analysis by highlighting different parts of speech, classifies text genres using color patterns to help students understand sentence structure~\cite{weber2007text}, teaches reading and composition through annotating different components~\cite{geigle2014color}, and identifies elements in analytical essays~\cite{olson2007cognitive}.
Notably, color-coding is particularly valuable in English as a Foreign Language education~\cite{azeez2021color}, as well as for students with learning disabilities~\cite{ewoldt2017color,maldonado2019color}.
%Despite ample research on color-coding for improved reading and comprehension, most have focused on adopting the technique from cognitive science or educational psychology.
%There is a notable lack of emphasis on quantitatively comparing color schemes, akin to usability studies in Human-Computer Interaction (HCI).
Most research on color-coding for better reading and comprehension comes from cognitive science or educational psychology, with little focus on quantitatively comparing color schemes in user tasks.
%, similar to usability studies in Human-Computer Interaction (HCI).

\subsection{Color in Visualization}
Previous studies explored color's role in visualization, particularly in areas like mapping quantitative data~\cite{tominski2008task}, categorical color maps~\cite{fang2016categorical}, or 3D spatial representations~\cite{besanccon2021state}.
The focus has primarily been on translating structural or numerical information into graphical forms. 
A notable example is the color scheme used in route tracing for the London metro map~\cite{lloyd2018metro}. 
Discussions on color choice in visualization predominantly center around graphical representations, such as investigating suitable color scales for spatial-temporal data~\cite{schulze2005enhancing}, color utilization in map-based information visualization~\cite{einakian2019examination}, recommendations for infographics color palettes~\cite{yuan2021infocolorizer}, and applying color palettes from historical artists' paintings into scientific image visualization~\cite{samsel2018art}.
%In our study, we examined how different color choices impact human perception when reading color-corded documents.
In contrast, textual information has received less attention.
Weber proposed a color-coding scheme for text highlighting based on English grammar parts of speech, relying on the author's interpretation of color meanings in Western culture~\cite{weber2007text}.
O'Connell and Fukao employed a reversed color-coded model, prompting students to match paragraph elements to colors based on context, with no explanation for the color choices provided in the research~\cite{o2019collaborative}.

\section{User Study}
%Thus, we propose a user study to explore the appropriate color palettes for annotating textual information to facilitate humans in information seeking. 
%This section overviews our user study.
%, each color-coded with a different scheme.  
%We conducted a pilot study with 3 types of color schemes borrowed from prior visualization studies of maps~\cite{harrower2003colorbrewer} and found that color schemes could lead to significantly different performances.

% \begin{figure*}[t]
%     \centering
%     \includegraphics[scale=0.5]%[width=\textwidth]{Figures/demo_interface/color-palette-interface-3.png}
%     {Figures/demo_interface/WebInterface-v3.png}
%     \vspace{-.5pc}
%     \caption{Web Interface Design with a timer and a question, using Arial font set at a 1rem size (equivalent to 16px)}
%     \vspace{-.5pc}
%     \label{fig:whole-ui}
% \end{figure*}

\subsection{Task Design and Interface}
We designed an information-seeking task that asks participants to identify the first text segment in the given abstract with a ``target label'' from our label scheme. 
The target label was one of the labels in the scheme, specified in the interface.
The task contains 123 question items: 3 trial question items and 120 main question items. 
Each question item consists of an abstract and a multiple-choice question: ``Which of the following is the FIRST segment belonging to the label [TARGET LABEL]?''
All four options were chosen from the first segment of each label in the annotated text to distract participants if they were focusing on the first segment for the wrong label. 
An 8-second time limit was set for each question item to ensure a controlled environment for task difficulty across participants.
Fig.~\ref{fig:color-palette-brief} shows our interface.
This task required participants to understand the matching between color and the label to seek information effectively.
The task workflow aligns with regular reading habits for English text and the way humans seek information from passages (top to down, left to right): 
1) Start by understanding what information to seek by identifying the target label (\eg, ``Method''), 
2) Check the color of the corresponding label (\eg, ``Method'' is indicated by ``Blue''), 
3) Locate the ``color block'', and 
4) Identify the first segment (top of the color block). 

The interface recorded the participant's answers and timestamps; we measured the response time on each question item by noting the start time at the commencement of the countdown timer.
The end time was marked at the point of the participant's final selection made within the 8-second frame.
In cases where the participant made no selection, no answer was recorded and the response time was automatically set to 8 seconds.
The results were exported as a JSON file after the task was completed for each participant. 
%of each session.

%\input{Figures/demo-figures}

\subsection{Experiment Material Preparation}
%\subsection{Color Schemes, Label Scheme, and Data}
%: Label Scheme, Color Schemes, and Data

%\kenneth{I made color selection and data processing subsubsections instead of paragraphs because they're kinda important or this paper.} \sam{It makes sense. Thanks!}

%\noindent\textbf{Color Schemes Selection.~}
\subsubsection{Color Schemes Selection}
Inspired by previous studies on the categorical perception of graphical stimulus using warm (red, yellow) and cool colors (green, blue) based on the CIELAB color space~\cite{holmes2017categorical}, we expanded this research to examine textual reading and a wider combination of colors in our study. % Explain the choice of color (Sam-0714-#4)
We adopted the Material UI color system\footnote{Material UI: https://mui.com/material-ui/customization/color/} and selected shade ``500'' for four identified hues: \textit{Red 500 (\#f44336)}, \textit{Yellow 500 (\#ffeb3b)}, \textit{Blue 500 (\#2196f3)}, and \textit{Green 500 (\#4caf50)} as the base color to generate the color schemes.
%\kenneth{Should we use this format: (Red: 0, Yellow: 60, Green: 120, Blue: 240)? See below for my next comment.} \sam{I used italic for color naming from Google and included the color hexcode to specify the color}
%\sam{make the text itatlic and add the hexcode for the color to make it more clear}
We used Colorgorical~\cite{gramazio2016colorgorical} to generate 10 different color schemes using all possible combinations of 4 base colors (Fig.~\ref{fig:color-palette-brief}), with several groups of color schemes: \textbf{Monochromatic} schemes utilize a single base color, while \textbf{Dichromatic} schemes employ two. 
\textbf{Analogous} schemes comprise colors of the same or adjacent hues on the color wheel, and \textbf{non-analogous} schemes incorporate colors from disparate hue ranges. %(Sam-0718-#13)
%\kenneth{Should we use the past tense here? Not entirely sure. (I mean, it's a factual description rather than stating a past action, so I don't know.) Sam?} \sam{I think present tense is appropriate as here is the definition of the scheme: factual description BUT for the verb "used" is in past tense because these are what we've done}
%We categorized the color schemes into three groups: warm, cool, and mixed. `Warm' includes schemes composed solely of warm colors, `cool' consists of schemes with only cool colors, and `mixed' contains a combination of both warm and cool colors.
%: 4 single-hue, 2 multi-hue (consistent), and 4 multi-hue (contrast). Figure~\ref{fig:demo_articles} shows the color schemes used in the main study.
We used the consistent setting for the scheme generation: 1) High in Perceptual Distance and Pair Preference, 2) Low in Name Difference and Name Uniqueness, and 3) lightness range: 50 - 90. For the hue filters, we set ``+/- 15\degree'' for each base color (Red: 0\degree, Yellow: 60\degree, Green: 120\degree, Blue: 240\degree), except ``+/- 35\degree'' for blue to generate effective color schemes.
%\kenneth{If we decide to use this format: (Red: 0, Yellow: 60, Green: 120, Blue: 240), be consistent throughout the paper.} \sam{I add the "degree" symbol to clarify the unit using}
We adopted the WCAG AA standard for normal texts that the contrast ratio is 4.5:1, using the contrast checker tools by WebAIM~\cite{wang2005webaim} to ensure sufficient contrast between the text color and each color from the generated color scheme.
If any color fails the test, another set of color schemes would be generated until all colors from the scheme pass the contrast test. 
%\footnote{WebAIM: https://webaim.org/resources/contrastchecker/}
%add the verb "would be generated" (Sam-0706-#3)
% correct "filer" to "filters" (Sam-0706-#4)
%\kenneth{Convert the two footnotes, Material UI and WebAIM, into citations. make sure the URL is still in the reference. Save some more space.}

\begin{table}[t]
\caption{Contrast ratios for 10 color schemes. The color was used as the baseline with the lowest contrast ratio is denoted by $\spadesuit$.} %$\medblackcircle$.}
\label{tab:contrast_ratios}
\centering
\small % Reduces font size to fit table on page
\begin{tabular}{|c|c|c|c|c|c|}
\hline
\textbf{Color Scheme} & \textbf{Color 1} & \textbf{Color 2} & \textbf{Color 3} & \textbf{Color 4} & \textbf{Color 5}  \\
\hline
Re & 5.56:1 & 5.55:1 & 5.55:1 & 5.56:1 & 5.56:1 \\
\hline
Ye & 5.63:1 & 5.57:1 & 5.55:1 & 5.59:1 & 5.57:1 \\
\hline
Gr & 5.57:1 & 5.6:1 & 5.56:1 & 5.6:1 & 5.59:1 \\
\hline
Bl & 5.61:1 & 5.59:1 & 5.6:1 & 5.63:1 & 5.56:1 \\
\hline
Re-Ye & 5.56:1 & 5.63:1 & 5.56:1 & 5.56:1 & 5.58:1 \\
\hline
Gr-Bl & 5.57:1 & 5.55:1 & 5.61:1 & 5.61:1 & 5.55:1 $\spadesuit$ \\
\hline
Gr-Re & 5.55:1 & 5.56:1 & 5.6:1 & 5.55:1 & 5.6:1 \\
\hline
Ye-Bl & 5.63:1 & 5.55:1 & 5.63:1 & 5.55:1 & 5.61:1 \\
\hline
Bl-Re & 5.55:1 & 5.56:1 & 5.61:1 & 5.57:1 & 5.61:1\\
\hline
Ye-Gr & 5.63:1 & 5.57:1 & 5.55:1 & 5.61:1 & 5.57:1\\
\hline
\end{tabular}
\vspace{-.5pc}
%\caption{Contrast ratios for 10 color schemes. The color was used as the baseline with the lowest contrast ratio is denoted by $\spadesuit$.} %$\medblackcircle$.} --> go above table
\vspace{-.5pc}
\end{table}

%$\largeblackcircle$

%Although black text is ordinary text for reading, different annotated colors with black text would create different contrast ratios. 
%, thus impacting the performance of locating information.
Our study focuses on the performance difference among the color schemes for annotation only. 
%However, it is unavoidable to exclude the combined effect of contrast between the highlight color and the text color.
%To minimize such contrast, we attempted to adjust the text color along the grey scale according to the highlight color by keeping the constant contrast ratio along all color schemes.  %(Sam-0715-#6)
%We considered using black for text which is common for reading, but combining with our selected annotated colors schemes would create different contrast ratios that might be critical factors to impact the performance. Thus, we adjusted the text color along the grey scale to maintain the constant contrast ratio for all colors in every color scheme. 
%To minimize effect caused by the different contrast between annotation color and text, we adjusted the text color along the grey scale to maintain the constant contrast ratio for all colors in every color scheme. 
We standardized contrast ratio to approximately 5.55:1 between highlight color and text color across all color schemes (Table~\ref{tab:contrast_ratios}).
%Table~\ref{tab:contrast_ratios} shows all the contrast ratios between each highlight color and text color. 
%To minimize systematic bias in visibility, we standardized the contrast ratio between each highlight color and its corresponding text color to approximately 5.55:1 across all color schemes by adjusting the text color along the grey scale (Table~\ref{tab:contrast_ratios}).\kenneth{Why was Table~\ref{tab:contrast_ratios} on Page 4 rather than here?? I moved it to here.} --> Sam: OK
This ratio was derived from the lowest acceptable contrast between the highlight color(\#658994 from Gr-Bl) and black text (\#000000).
%We used the lowest contrast ratio (5.55:1) among all the individual color from the generated scheme (\#658994 from Gr-Bl) when the text was fully black. %(Sam-0717-#12)
%For other colors with a higher contrast ratio, when the text is fully black, we adjust the text color to lighter until reaching the closest possible value that is at least the lowest contrast ratio. 

%\sam{@Steven: Please help to update the description here}\steven{done}
\subsubsection{Data Processing and Label Scheme}
The data used in the study contained abstracts extracted from recent publications in the arXiv HCI (\texttt{cs.HC}) field (from February to April 8th, 2024).
%This data, obtained on April 8th, 2024.
%was selected with the expectation that most participants would be graduate students from a similar field.
We aimed to simulate the information-seeking scenario of college students reading paper abstracts during the literature review process.
%In this study, 
We used CODA-19's label scheme~\cite{huang-etal-2020-coda}, categorizing sentence segments in paper abstracts into \textbf{Background, Purpose, Method, Finding, and Other}.
We chose this scheme because it is closely relevant and useful for a graduate student's reading scenario.
For instance, a student might use color-coding annotations to identify the ``Finding'' of a paper.
To process the data, we first used Stanford CoreNLP~\cite{manning-etal-2014-stanford} for tokenizing and segmenting sentences in all the abstracts. 
We then used commas (,) and periods (.) to divide sentences into smaller text segment parts.
Each segment contained at least six tokens (excluding punctuation), ensuring the first segment in each label had sufficient content.
We selected 706 abstracts with a token range from 150 to 250.
The average abstract had 7.28 sentences (SD=1.47), which were further divided into 12.42 text segments (SD=2.50).
Each abstract had 189.29 tokens (SD=26.32).
We then used He's~\cite{he2024chi} zero-shot prompt on OpenAI's GPT-4~\cite{openai2023gpt4}, which had a high accuracy (83.6\%) evaluated by expert labels, to automatically label each segment in abstracts for this study.
We excluded abstracts that lack one of four primary classes, namely Background, Purpose, Method, and Finding.
% Sam-0715-#8
%For the purpose of this study, we excluded abstracts that lack one of four primary classes, namely Background, Purpose, Method, and Finding.

% Finally, we excluded 15 abstracts lacking labels for all four primary classes, namely Background, Purpose, Method, and Finding.

% task --> 
% question --> abstract + question + 4 options
% abstract --> text from arxiv

%\sam{@Steven: Please help to update here too} \steven{done}
%\sam{I tried to modify the number here, Steven can check if correct}
\paragraph{Data Selection.}
From the labeled abstracts, we randomly sampled 123 unique abstracts for the user study.  % (Sam-0715-#5)
%based on the following criteria: 
%(1) each abstract had tokens ranged from 150 to 250 to ensure sufficient and consistency in the amount of textual information
%(2) each abstract contained at least 4 main labels (Background, Purpose, Method, Finding).
%, which will be explained further in the next paragraph. This is because "Other" is not available in many abstracts. 
The first 3 abstracts were used for trial question items. %\kenneth{Again, past tense! (I fixed it.)}
We divided the remaining 120 abstracts evenly among 10 color schemes for color annotation.
Within each color scheme, 12 abstracts were further divided into 4 main labels, with each label corresponding to 3 abstracts. This division created a set of question items for each %label-abstract pairing.
color scheme.

% For the 12 abstracts in each color scheme, they were assigned equally with 4 target labels (Background, Purpose, Methods, Finding) to compose the question set of each color palette.

\subsection{Study Procedure and Setups}

\paragraph{Participants.}
We recruited 32 participants for the main study, with diverse educational backgrounds (12 Master's students, 15 Ph.D. students, and 5 Undergraduate students).
All participants reported no visual impairment or color blindness.

\paragraph{Study Setups.}
The study took place in person in a university campus lab room, which featured a quiet environment and full white lighting to provide optimal conditions. 
Participants used a 24" BenQ EW2440L display monitor connected to a laptop (ROG Zephyrus G15) via HDMI cable. 
The monitor had a resolution of 1920 x 1080 with standard dynamic range (SDR) color space and specific settings: 1) Blue Light Setting: Low Blue Light - Multimedia (30\%), 2) Brightness: 60, 3) Contrast: 50, and 4) Sharpness: 6. 
The monitor was centrally positioned on the desk, with a distance of 21" from the display edge to the desk (approximately the distance between participants' eyes and the monitor). 
This setup and setting remained constant for all participants throughout the study.
During the task session, the researcher observed from a distance of approximately 40 inches to minimize disruption.

\paragraph{Study Procedure.}
Upon arrival, a researcher (one of the paper's co-authors) provided a briefing about the study's purpose, obtained oral consent, and guided the participants through a tutorial document to understand the web interface and tasks. 
The tutorial, displayed on the BenQ monitor in PDF format, ensured consistency between the tutorial and the actual task session.
After addressing participant questions, the researcher launched the web interface and started screen recording.
Participants entered their names and clicked ``start'' when ready.
Participants first encountered the 3 trial question items to ensure they understood the task before proceeding.
Then, they proceeded to complete the question set of 10 color schemes one by one, with 30 seconds in between each color scheme.
To minimize the order effects, we randomized the order of question items within each color scheme and also the sequences of color schemes presented to each participant.
%After completing the test, the researcher stopped the screen recording and directed the participants to complete an exit survey, which required the participants to look at the same passage of text for each color scheme one by one and rate their preference for each color scheme as shown in (Figure~\ref{fig:color-palette-brief}) using the 5-point Likert scale. % (Sam-0715-#7)
After completing the test, participants completed an exit survey, rating their preference for each color scheme on a 5-point Likert scale, using the same text passage for all schemes (Fig.~\ref{fig:color-palette-brief}). % (Sam-0715-#7)
The researcher then conducted follow-up interviews based on test observations and participants' preferences.
Upon completion of the study, participants received a \$5 cash compensation, confirmed by signing a compensation form.
The user study took approximately 25 minutes for each participant.

\section{Findings\label{sec-finding}}

\begin{figure}[t]
    \centering
    \includegraphics[width=1\linewidth]{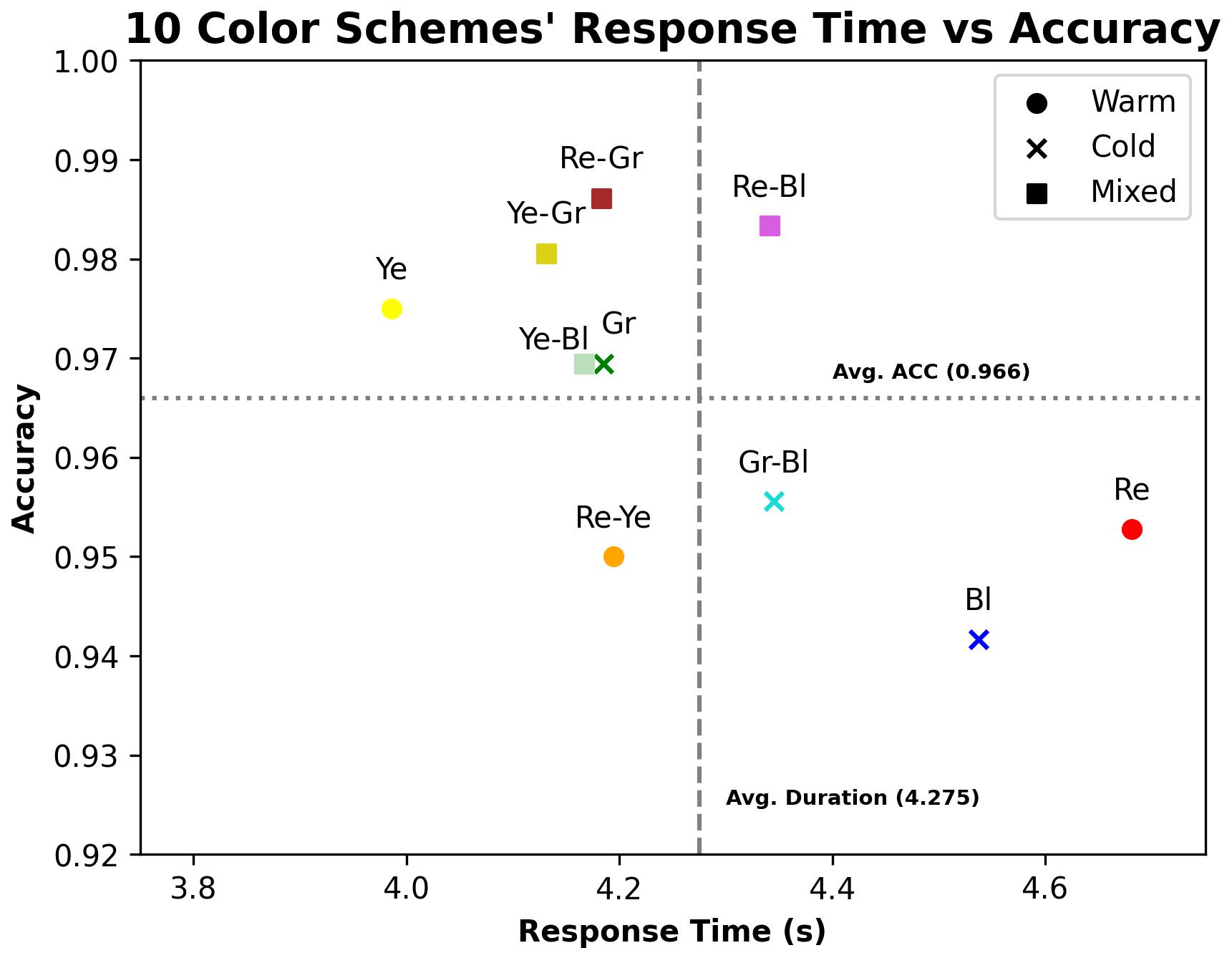}
    \vspace{-.5pc}
    %\vspace{-1.5pc}
    \caption{Response Time and Accuracy Plot for 10 Color Scheme. Each color scheme was classified into 'Warm,' 'Cool,' and 'Mixed' categories and was denoted by distinct symbols.}
    \vspace{-.5pc}
    \label{fig:dur-acc-plot}
        %\kenneth{TODOs: (1) Just white background with NO grid lines, (2) add error bar}\steven{done. error bar is using 95\% confidence interval}

\end{figure}

\begin{table}[t]
    \caption{User study results on different combinations of color palette. Accuracy pairs and Response Time pairs that passed the T-Test are denoted by 
    $\dagger$ (p-value = 0.001), 
    $\diamondsuit$ (p-value = 0.020),
    $\heartsuit$ (p-value = 0.029),
    $\triangle$ (p-value = 0.024),
    and $\clubsuit$ (p-value $<$ 0.001).
    % P-value in the table are only for comparison between Diverging and Sequential and between Qualitative and Sequential. \steven{will add this later}
    }
    \label{tab:color-whole-table}
    \centering
    \small
    \begin{tabular}{ccccccccccc}
    \toprule
    \multirow{2}{*}{\textbf{n=30}} &\multicolumn{2}{c}{\textbf{ACC}} &\multicolumn{2}{c}{\textbf{Response Time}} \\ 
    \cmidrule(lr){2-3} \cmidrule(lr){4-5}
    \multicolumn{1}{c}{}&\multicolumn{1}{c}{Mean}  &95\% CI  &\multicolumn{1}{c}{Mean}  &95\% CI  \\ \midrule
    \multicolumn{1}{c}{Monochromatic} &.960 &{[}.938,.981{]} &4.348$^\diamondsuit$&{[}4.163,4.532{]} \\		
    \multicolumn{1}{c}{Dichromatic} &.971 &{[}.956,.985{]} &4.227$^\diamondsuit$&{[}4.067,4.387{]}\\ \midrule 		
    \multicolumn{1}{c}{Analogous} &.957$^\dagger$ &{[}.939,.976{]} &4.322$^\heartsuit$&{[}4.160,4.483{]} \\		
    \multicolumn{1}{c}{Non-Analogous} &.980$^\dagger$ &{[}.966,.993{]} &4.206$^\heartsuit$&{[}4.022,4.389{]}\\ \midrule   		
    \multicolumn{1}{c}{Cool Only} &.956&{[}.933,.978{]} &4.356&{[}4.190,4.522{]}\\ 		
    \multicolumn{1}{c}{Warm Only} &.959&{[}.936,.983{]} &4.287&{[}4.113,4.461{]}\\ \midrule		 		
    \multicolumn{1}{c}{Red-Inclusive} &.968&{[}.950,.986{]} &4.350$^\triangle$&{[}4.179,4.521{]}\\ 				
    \multicolumn{1}{c}{Red-Exclusive} &.965&{[}.946,.985{]} &4.225$^\triangle$ &{[}4.054,4.396{]}\\ \midrule				
    \multicolumn{1}{c}{Yellow-Inclusive} &.969&{[}.949,.989{]} &4.120$\clubsuit$&{[}3.950,4.289{]}\\ 				
    \multicolumn{1}{c}{Yellow-Exclusive} &.965&{[}.950,.980{]} &4.379$\clubsuit$&{[}4.212,4.545{]}\\ 	\midrule			
    \multicolumn{1}{c}{Green-Inclusive} &.973&{[}.961,.985{]} &4.211&{[}4.037,4.385{]}\\ 				
    \multicolumn{1}{c}{Green-Exclusive} &.962&{[}.939,.985{]} &4.318&{[}4.128,4.508{]}\\ 	\midrule			
    \multicolumn{1}{c}{Blue-Inclusive} &.963&{[}.936,.989{]} &4.348&{[}4.147,4.549{]}\\ 			
    \multicolumn{1}{c}{Blue-Exclusive} &.969&{[}.954,.984{]} &4.227&{[}4.061,4.392{]}\\ 		
    % \midrule
    % \textbf{Avg.} &.967 &-&.878&-&5.25&-\\ 
    \bottomrule

    \end{tabular}
    \vspace{-.5pc}
    %\caption{User study results on different combinations of color palette. Accuracy pairs and Response Time pairs that passed the T-Test are denoted by 
    %$\dagger$ (p-value = 0.001), 
    %$\diamondsuit$ (p-value = 0.020),
    %$\heartsuit$ (p-value = 0.029),
    %$\triangle$ (p-value = 0.024),
    %and $\clubsuit$ (p-value $<$ 0.001).
    % P-value in the table are only for comparison between Diverging and Sequential and between Qualitative and Sequential. \steven{will add this later}
    %}
    %\vspace{-1pc}
    \vspace{-.5pc}
\end{table}

%There was no significant different between Diverging and Qualitative on Response Rate, ACC, and Duration.

%\kenneth{TODO: (1) Add t-test results to them. (2) We don't really need a row for Avg I think.}\steven{done}

% \input{Figures/bar-chart-figures}
% \sam{TODO: need update with latest data} \steven{done}
% From the task session, we calculated accuracy and response time. 
% Regarding accuracy, we considered attempts where participants failed to submit their answers within 8 seconds incorrect. 
% For response time, we used 8 seconds for attempts in which participants failed to submit their answers.
%\label{sec-finding}
Before analysis, we cleaned the data by removing two outliers who had low accuracies (0.25, 0.65) that fall below the average minus two standard deviations (0.934 - 2$\times$0.140 = 0.654)~\cite{ratcliff1993methods}.
The new accuracy after pruning is 0.966 (SD = 0.041).
We documented the two outliers' behavior for detailed discussion in Section~\ref{sec-discussion}.
%Duration mean 4.275s and  0.429 std
%response rate 0.988 0.0231

% \sam{TODO: update with 10 color instead}
\paragraph{Participants performed significantly better when using non-analogous color schemes than analogous color schemes.}
Fig.~\ref{fig:dur-acc-plot} shows the results for all individual color schemes.
%with the x-axis representing the response time and y-axis indicating accuracy.
% For time, yellow scheme had the shortest time and Red scheme had the longest time time. Red scheme was significantly slower than all color scheme except Blue scheme. Yellow was only significantly faster than half of the schemes. 
% As of accuracy, Red-Green scheme had the highest accuracy and Blue scheme had the lowest accuracy. 
%Though Red-Green had the highest accuracy, it was only significantly higher than Green-Blue and Red schemes. Interestingly, schemes whose accuracy that significantly higher than Red scheme were mixed schemes. 
%In the figure, each colored dot represents the results of each color scheme in terms of accuracy and response time. 
% We categorized the color schemes into three groups: warm, cool, and mixed. `Warm' includes schemes composed solely of warm colors, `cool' consists of schemes with only cool colors, and `mixed' contains a combination of both warm and cool colors. These categories are denoted by distinct symbols.
% \steven{should we mentioned about the t-test between each individual color?}
Interestingly, the accuracies of all `mixed' color schemes were above average, with most also achieving faster-than-average response times. 
In contrast, the majority of `cool' color schemes displayed both accuracies and response times below average.
The results for `warm' color schemes varied, with their accuracies and response times distributed throughout the figure.

%\kenneth{``it is surprising that red and yellow are both warm color groups but lead to two different direction effects in response time: response time for color schemes generated by red as base color leads to a significantly slower response, while yellow leads to a significantly faster response.
%It echoes other psychological research about "yellow priority" that yellow is more salient compared to other colors.''---> This can go in Findings. Do not put in Discussion.}

Given that color schemes with mixed colors performed noticeably better than those with warm/cool colors, we sought to explore the factors influencing accuracy and response times. 
%To achieve this, 
We set up several pairwise comparisons to clarify the factors affecting performance. These comparisons included monochromatic versus dichromatic schemes, analogous versus non-analogous schemes, warm colors versus cool colors, and color-inclusive versus color-exclusive schemes.
%More specifically, the monochromatic schemes are schemes generated by a single base color, while the dichromatic schemes are generated by two base colors. We defined analogous schemes as those generated solely by warm color(s) or solely by cool color(s), whereas non-analogous schemes were generated by warm-cool color pairs. For color-inclusive versus color-exclusive schemes, we evaluated schemes generated by a specific base color against schemes that excluded that base color. 
%Table~\ref{tab:color-whole-table} shows the results for different combinations of color schemes. Participants exhibited significantly better performance when using non-analogous color scheme to locate information. The accuracy and response time of analogous and non-analogous schemes show statistical differences. In comparison to cases using the analogous color schemes, participants had higher accuracy (p = 0.001) and spent less time (p = 0.029). Meanwhile, dichromatic schemes demonstrated a significantly faster response time than monochromatic schemes. Within color-inclusive versus color-exclusive schemes, red-inclusive schemes resulted in a significantly slower response than red-exclusive schemes, while yellow-inclusive schemes resulted in a significantly faster response than yellow-exclusive schemes.
Table~\ref{tab:color-whole-table} shows the results for various color scheme combinations. Non-analogous schemes significantly improved participants' information location performance, showing higher accuracy (p=0.001) and faster response times (p=0.029) compared to analogous schemes. 
Dichromatic schemes yielded significantly faster responses than monochromatic ones. 
Among color-specific schemes, red-inclusive schemes slowed responses, while yellow-inclusive schemes accelerated them, compared to their respective color-exclusive counterparts.
% Participants exhibited significantly poorer performance when using the \textbf{sequential} color scheme to locate information.
% In comparison to cases using the other two color schemes, participants responded less (p<0.013, compared with \textbf{qualitative}), with lower accuracy (p<0.001, compared with both), and spent more time (p<0.001, compared with both).

% Meanwhile, the \textbf{diverging} and \textbf{qualitative} color schemes showed similar performance. 
% It is important to note that our label scheme, based on CODA-19~\cite{huang-etal-2020-coda}, is closer to a qualitative scheme. 
% However, the \textbf{diverging} color scheme actually resulted in slightly higher labeling accuracy and slightly shorter task time than the qualitative color scheme; the difference was just not statistically significant.

% \input{Figures/user-rating}

\paragraph{Participants preferred yellow-inclusive schemes and disliked red-inclusive schemes.}
% About preference, the color schemes contain yellow as base color are preferred by most participants, showing by the frequencies of highest rating from all participants, including Ye-Gr (18), Ye-Bl (16) and Ye (15) respectively. 
In the exit survey, participants were asked to rate the color schemes with the following statement, ``This color palette helps me to recognize the structure of the abstract at glance.'' for each color scheme using a 5-point Likert scale from Strongly Disagree to Strongly Agree. The three most preferred color schemes based on were Ye-Gr (4.581), Ye (4.226), and Ye-Bl (4.194). Conversely, the three least-liked color schemes were Re (2.419), Bl-Re (2.968), and Gr (3.387).

In addition to quantitative data, we gathered qualitative insights through observation and brief interviews with participants. 
For yellow-inclusive color schemes, which had comparably better performance in response time and better preference, 
many participants found it effective for quickly identifying information.
One participant mentioned that ``yellow and light colors are the best''. 
For red-inclusive color schemes, which resulted in poor performance and preference, many participants mentioned that ``the red colors are too bright and distracting.'' 
For cool color schemes, most participants did not give comments, except one participant showed a strong preference towards green: ``I find Green as an easy-to-go color, other colors are creating an information overload in the mind.''
%\kenneth{This quote is not very fluent. Is it a translation? If so, add [sic] into the quote to make it clear: https://www.grammarly.com/blog/sic/; if not, use a better translation.} \sam{attempted to correct the translation}

\section{Discussion\label{sec-discussion}}
\paragraph{The impacts of red and yellow.}
%\paragraph{The impact of specific colors towards information-seeking in textual documents.}
%While the result might not show there are specific impacts of information-seeking performance due to the color temperature, especially the large deviation between the performance between red and yellow, which both are warm colors, it gives us some insights into how specific color or the combination of color could lead to different performance. 
%First, the color schemes with mixed color temperature gives very significantly higher accuracy than other settings (p$<$0.001). It may be due to the higher contrast among colors than in the homogeneous color schemes, which seems to be consistent with most of the previous research about categorical identification that higher contrast between colors helps participants to identify targets correctly.
%Second, 
It is surprising that red and yellow are both warm color groups but led to two different direction effects in response time: response time for color schemes generated by red as base color led to a significantly slower response, while yellow led to a significantly faster response.
It might echo other psychological research about the ``yellow priority'' that yellow is more prominent compared to other colors~\cite{hu2020yellow}. 
More research is needed to understand the causes.
%This could also explain the majority preference towards the color schemes with yellow.

\paragraph{Suggestions for color coding of textual documents.}
A few suggestions emerged from our findings that can be used to inform the color-coding practices of textual documents:
%Our findings show that the knowledge and materials employed in studying colors for graphical representations can be used to guide the study of color coding of textual documents; 
1) Color schemes with \textbf{mixed color temperatures} give better performance and are recommended to be used for annotating textual documents to help readers accurately search for the information from specific labels.
2) \textbf{Yellow is recommended} to be included in the color schemes to help the reader locate information in a shorter time. 
3) \textbf{Avoid the red} in the color schemes as it could delay the search time and cause discomfort to readers' eyes.
%These suggestions could inform LLM prompts for text color-coding.% (Sam-0715-#10)
These suggestions could inform the selection of color schemes for color-coded documents.
% (Sam-0715-#10)
%\kenneth{What does "inform LLM prompts" mean? Just say ``the selection of color schemes for color-coded documents''?} \sam{Because I think the reviewer thinking about this paper is more related to the LLM, so try to incorporate the idea of "our paper try to inform the color selection for doing color coding using LLM" --> but you are right that the underlying concept is indeed "color schemes for color-coded documents"}

\paragraph{Outliers' Behavior.}
Two participants were identified as outliers in the study due to specific circumstances affecting their performance. 
One participant reported experiencing significant stress and frustration during the task, citing their slow reading speed and insufficient time to process the text options before selecting answers. 
This highlights the impact of cognitive load and the need for inclusive design considerations in tasks involving reading. 
The other participant expressed discomfort using the mouse, which impeded their ability to select answers quickly in several questions. 
These cases show a potential limitation in the study design, specifically the requirement for participants to use a mouse to complete the task, which may not accommodate all users equally.

\paragraph{Limitations.}
We acknowledge a few limitations in this work. 
Firstly, our study focused on examining the impact of color schemes on locating specific textual information, omitting a direct assessment of reader comprehension.
While comprehension is crucial, it is more resource-intensive to evaluate than information location. 
We identify this as a potential avenue for future research.
%Secondly, readers with a non-native English background might encounter difficulties reading the text. 
%Although many of our participants were non-native English speakers, they were graduate students studying in the U.S. with generally proficient English fluency.
%Our findings may not be readily applicable to individuals with lower English proficiency.
Secondly, perceived visibility or ``pop-out'' effect may still vary across different color combinations due to factors such as chromatic contrast and individual perceptual differences~\cite{wool2015salience} despite our effort to standardize the contrast ratio between each highlight color and the respective text color.
%\kenneth{What's a pop up effect?? Can we cite something here?} \sam{pop-up effect is mentioned by the reviewer, and it means that some colors in nature are more attractive and more recognized by human eyes --> even though we tried  to control the same contrast between text/high light color, but some color catch more human's attention in nature (e.g. red) --> I found a paper talking about this (should be "pop-out" effect) and cited it}
Future studies could benefit from incorporating more advanced color appearance models or conducting supplementary perceptual tests to address color visibility.
%Lastly, our study exclusively focused on scholarly articles, a highly specialized genre not tailored for a general audience.
%The generalizability of our results could be further explored by conducting similar studies on other types of textual information.
Lastly, our focus on scholarly articles limits generalizability. Future studies should examine these effects across diverse text types.

\section{Conclusion and Future Work}
This paper conducts a user study to assess the impact of 10 color schemes, generated from the combinations of 4 base colors, on rapid and accurate information seeking in color-coded documents. 
%Drawing inspiration from prior visualization studies for maps, combined with the existing color theory about the color temperature of cool and warm to generate 10 color schemes. 
%We picked the 10 color schemes, featuring a combination of different categories, %are inspired by prior visualization studies for maps and existing color theory on the color temperature of cool and warm.
%based on the color temperature of cool and warm. %\kenneth{This sentence is strange. 1) Past tense, 2) in the conclusion you can probably just say 10 color schemes without explaining how we picked them.} --> Sam: Make it more simplier and incorporate into the first sentence
% (Sam-0715-#8)
The results indicate that the \textbf{non-analogous} color scheme leads to better information-seeking performance, \textbf{yellow-inclusive} schemes lead to shorter response time and are also more preferred by most participants. 
These could inform the better choice of color scheme for annotating text documents.
%The study underscores the influence of color scheme selection on readers' performance in seeking information in color-coded documents.
%This study encourages discussions and investigations into color coding for textual information. 
As LLMs enhance our ability to code textual documents, we advocate for additional research specifically focusing on the ``color'' aspect of color-coding techniques.

% \appendix
% \input{sections/7_appendix}

%\acknowledgments{
%The authors wish to thank A, B, and C. This work was supported in part by
%a grant from XYZ.}

%\bibliographystyle{abbrv}
\bibliographystyle{abbrv-doi}

\bibliography{bib/ColorRef}
\end{document}